# Modeling and Simulation of the EV Charging in a Residential Distribution Power Grid

Fereidoun Ahourai, Irvin Huang, and Mohammad Abdullah Al Faruque
Department of Electrical Engineering and Computer Science
University of California, Irvine
Irvine, California, USA
{fahourai, ibhuang, alfaruqu} @ uci.edu

***Abstract*—There are numerous advantages of using *Electric Vehicles* (EVs) as an alternative method of transportation. However, an increase in EV usage in the existing residential distribution grid poses problems such as overloading the existing infrastructure. In this paper, we have modeled and simulated a residential distribution grid in GridLAB-D (an open-source software tool used to model, simulate, and analyze power distribution systems) to illustrate the problems associated with a higher EV market penetration rates in the residential domain. Power grid upgrades or control algorithms at the transformer level are required to overcome issues such as transformer overloading. We demonstrate the method of coordinating EV charging in a residential distribution grid so as to overcome the overloading problem without any upgrades in the distribution grid.**

## I. INTRODUCTION AND RELATED WORK

There is a growing trend towards *Electric Vehicles* (EVs) as a medium of transportation because of their economic and environmental benefits [1]. The EV is not a new concept and has been conceptually and practically available for the last century since the advent of automobiles [3]. Today the cost of electricity to drive the EV is becoming competitive with the cost of fossil fuel required to drive the same distance [2][4]. The price of an EV is still higher than its traditional counterpart running on fossil fuel but the total cost is lower in the long term due to less maintenance being needed for EVs and overall price spent on fuel per mile [4]. EVs also reduce the consumption of natural fossil fuels and make the environment clear by reducing the *Green House Gas* (GHG) emissions [5].

Higher rates of EV market penetration will have a negative impact on the electric power grid because uncoordinated EV charging on a mass scale at the secondary distribution grid would negatively affect the total load and peak load power [7][8][10]. The results in [7] show that 30% of peak load power usage is expected due to EV charging in the distribution grid. To address this negative affect, one proposal is to upgrade the electricity infrastructure by changing the transformers and adding more power plants to provide more energy to the residential grid [5][8]. This solution will undermine the economical and environmental benefits of EVs. Another solution is to control and coordinate the EV charging locally and at the substation level to mitigate the impact of EV charging from generating the peak power [6][10].

In [10], authors use a demand response model (time-of-use) strategy to shape the peak load of EV charging in the smart grid. Authors in [7] delayed the EV charging schedule to reduce the peak load. A co-simulation of OMNeT++ and OpenDSS has been conducted to show the impact of data rate-based and event-based models on the control algorithms of EV to reduce the peak load in [9]. However, none of the previous approaches have used an accurate model of the residential distribution grid with *Electric Vehicle Supply Equipment* (EVSE) that may provide variable amount of current on-demand to EV and EV departure and arrival model.

Therefore, the scope of this paper is to model the EV and a residential distribution gird in GridLAB-D to illustrate the effects of the different EV penetration rates in the power grid in order to design and validate a control algorithm for coordinating EV charging. GridLAB-D is an open-source power system modeling and simulation tool developed by *Pacific Northwest National Laboratory* (PNNL) with the funding of the *Department of Energy* (DOE) [13][14]. GridLAB-D is a discrete event-based power systems simulator which employs an agent-based simulation approach to model and simulate the distribution power grid [14]. Existing EV model in the GridLAB-D is not complete; therefore, we have modeled an advanced EV model to develop the collaborative charging algorithms[1] in a residential distribution grid.

The rest of the paper is organized as follows: in section II, we present our novel contributions. Section III describes our residential distribution grid model. Our EV charging algorithm is presented in section IV. Section V details the simulation results, and finally section VI concludes the paper.

## II. OUR NOVEL CONTRIBUTIONS

We have modeled the residential distribution grid to simulate its various dynamic properties under various EV penetration rates. Our contributions within the scope of this paper are as follows:

- We have modeled a residential distribution grid using the power system modeling tool GridLAB-D.

[1] Coordinates EV charging among multiple neighboring households collaboratively

- We have implemented advanced EV and EVSE models within GridLAB-D for our presented distribution grid.
- We have demonstrated the impact of EV charging under various penetration rates for our developed model.
- We have demonstrated an EV charging coordination algorithm for the residential distribution grid.

## III. Residential Distribution Grid Model

In this section, we describe our residential distribution grid model and the parameters of the objects. Moreover, we explain the EV and EVSE models used in our residential distribution grid. Our model consists of two parts: structural and behavioral. In structural modeling, we explain the structure of the distribution grid while in behavioral modeling, we explain the dynamic parts of the power system. In [11], we have described our model in detail.

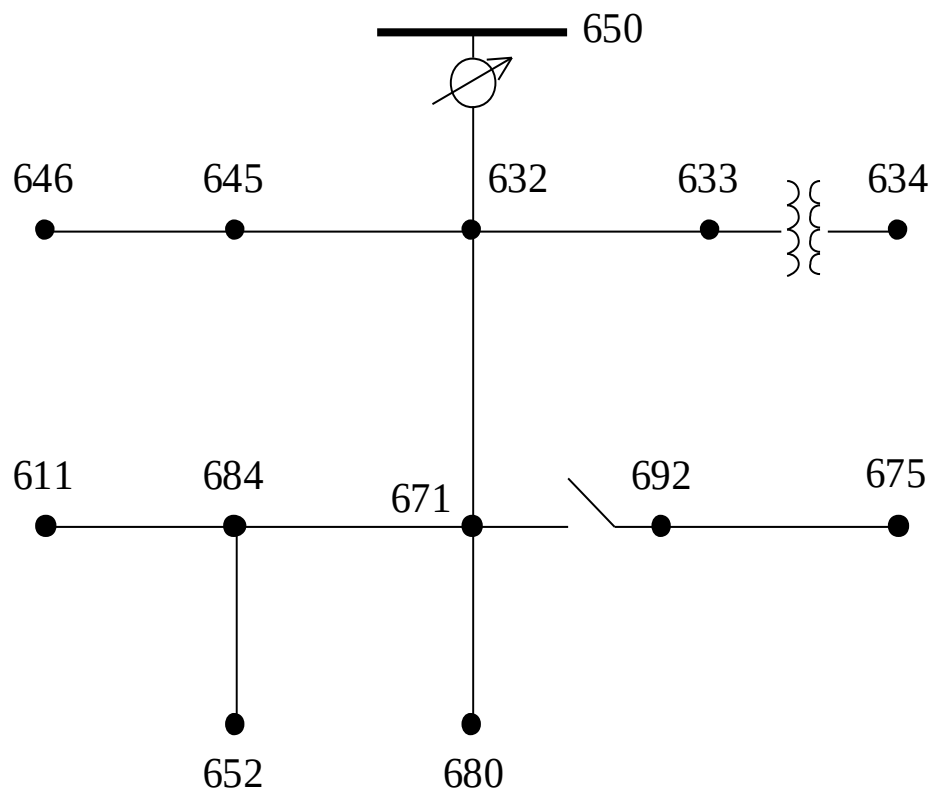

Fig. 1. IEEE 13 Node Test Feeder [12]

### A. *Structural model*

The structural model of our residential distribution grid consists of the distribution feeder, step down transformers, a triplex meter, two types of houses, and a variety of appliances. We have modeled the structure of the residential distribution grid with our modeling tool GridLAB-D by using IEEE 13 node [12]. Figure 1 shows the single line diagram of the IEEE 13 Test Feeder. The primary voltage of feeder is 33KV and the secondary voltage is 2.4KV. The power rate for the substation transformer is 5MVA. Each step-down transformer connects to one of the IEEE 13 nodes. The primary voltage for this type of transformer is 2.4KV while the secondary voltage is 120V. For the step-down transformer power rate, we have assumed that the transformer rate is 5KVA for every house which is connected to a transformer.

We have randomly distributed 1000 houses among the nodes of the IEEE 13 node feeder in which every 3 to 7 houses connect to a step-down transformer. We have characterized two types of single-family houses in our residential distribution grid model. *Type 1* houses represent houses with lower power consumption while *Type 2* houses represent bigger houses with higher power consumption. The houses we have used are randomly selected from the two types, both of which have different end-use appliances such as dishwasher, lights, water heater, plug load (miscellaneous), refrigerator, clothes washer, dryer, and oven. Table I shows the physical model and appliance specifications for each type of houses. We have extracted this information from the most up-to-date sources such as the U.S. DOE [15], OkSolar [16], and [7].

TABLE I. House Model and Appliance Specifications

| | Type 1 | Type 2 |
|---|---|---|
| Number of stories | 1 | 2 |
| Floor area | 2100 sq. ft. | 2500 sq. ft. |
| Heating system | GAS | GAS |
| Cooling system | Electric | Electric |
| Thermal integrity | Normal | ABOVE_AVERAGE |
| Motor efficiency | AVERAGE | AVERAGE |
| Number of occupants | 3 | 5 |
| Heating set point | 68 F | 68 F |
| Cooling set point | 72 F | 72 F |
| Light power | 1.2kW | 1.5 kW |
| Dishwasher power | 1 kW | 1.5 kW |
| Water tank volume | 40 gal | 50 gal |
| Water heater power | 3 kW | 4 kW |
| Clothes washer power | 0.8 kW | 1 kW |
| Miscellaneous | 0.7 kW | 0.8kW |
| Compressor power | 0.5kW | 0.6kW |
| Oven | 2.4kW | 3kW |
| Oven set point | 500 F | 500 F |
| Dryer | 2kW | 3kW |

To model the impact of EV penetration in our residential distribution grid, we have considered five different penetration rates of EVs in the residential domain (0%, 10%, 20%, 30%, and 50%). In this model, we have assumed that all the EVs arrive at their households following the Gaussian probability distribution model in which the mean is 5:30 PM and the standard deviation is 1 hour. The battery size is either 25kWh or 40kWh according to Table II. We have used the same assumption for determining when the EVs leave their households (EV departure follows a Gaussian probability distribution with mean equals to 7:30 AM). Figure 2 illustrates the arrival and departure time of the EV.

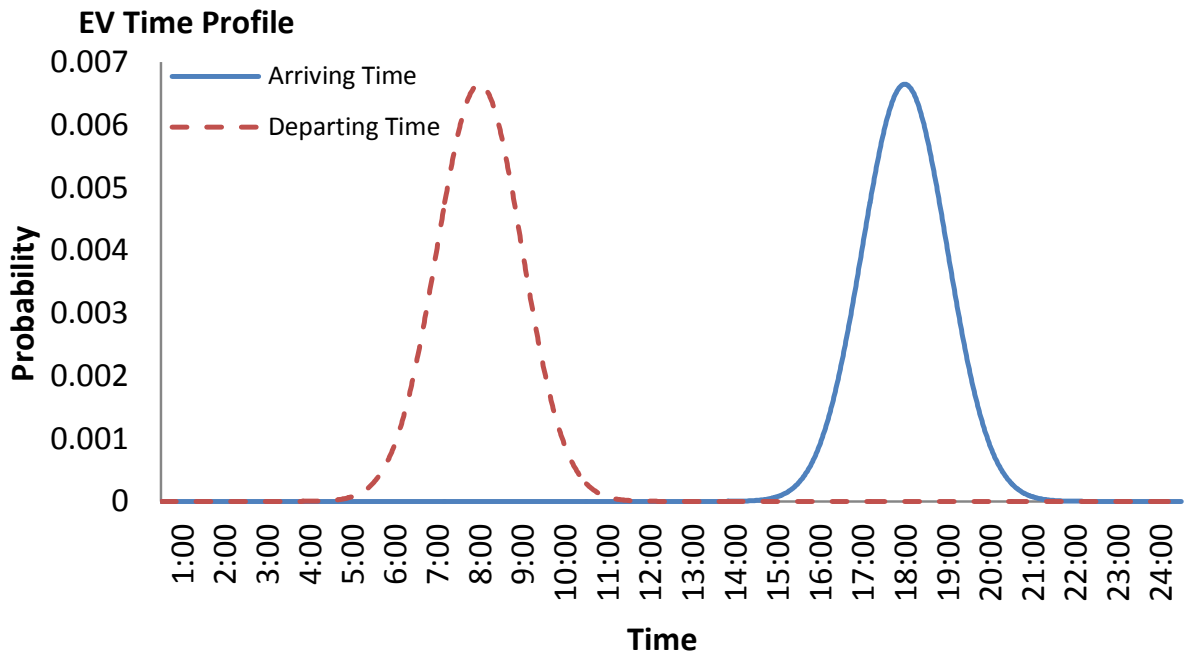

Fig. 2. Departure and Arrival Time Profile for the EV

We have also used the Gaussian probability distribution model for the total distance the EV drives every day. When the EV arrives home, the *State Of the Charge* (SOC) is modeled using the Gaussian probability distribution theory with mean value according to Table II.

TABLE II. EV Model Specification

| House | EV Battery Size | Miles Classification | State of Charge | Charging Amp. | Charging Volt. |
|---|---|---|---|---|---|
| Type 1 | 25kWh | 75 miles | 20% | 30A | 240V |
| Type 2 | 40kWh | 140 miles | 25% | 30A | 240V |

In our model, each EV has one EVSE which is installed inside each household. The EVSE starts to charge the EV as soon as it arrives home and stops charging when the battery is full or the EV leaves. When there is no charging strategies, the ESVE provides a constant charge at 30A (the maximum current), and the battery gets charged at a rate of 7.2 kWh. The EVSE gets some information from the EV model such as SOC, time and distance of the next trip, mileage classification, and battery size. These parameters can be used to control the charging rate of the EV through the EVSE[2] when we want to implement EV charging algorithms. Figure 3 shows the different states of EV. Equation 1 shows the EV battery level during charging and Equation 2 shows the EV battery level during discharging. $SOC$, $b_c$, $E$, $c_e$, $s$, $d$, and $m_e$ are state of charge, battery capacity (amount of charge in KWh), energy charging rate, charging efficiency, battery size, distance, and mileage efficiency, respectively.

$$SOC_{Charge} = \frac{b_c + E \times c_e}{s} \times 100 \qquad (1)$$

$$SOC_{Discharge} = \frac{b_c - \dfrac{d}{2 \times m_e}}{s} \times 100 \qquad (2)$$

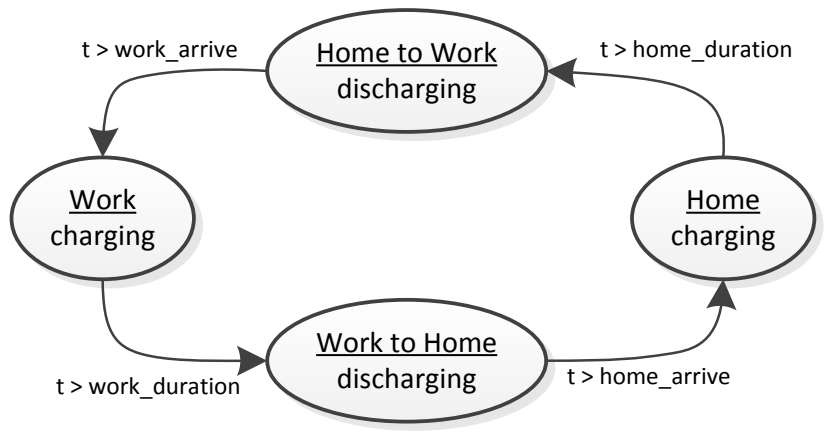

Fig. 3. EV Model State Chart

## B. Behavioral model

Each appliance in our model has a schedule which is attached to it. The schedules are assumed to be randomly distributed with a variance proportional to the mean. Figures 4 and 5 show the schedules for lights and water heaters, respectively. For the weather, we use the city of Newark, New Jersey.

# IV. EV Charging Controller

To demonstrate the effect of coordinated EV charging in a residential distribution grid, we have developed the fair sharing algorithm to mitigate the peak load in the residential distribution grid. The fair sharing algorithm monitors the power output of step-down transformers every minute and equally divides the remaining output power among all EVs connected to this transformer.

$$Charging_{rate} = \frac{T_r - (T_o - \sum EV_r)}{n} \qquad (3)$$

The algorithm begins by sending a signal to the EVSE which changes the charging current (amperage) between 0 to 30A. Equation 3 shows the charging rate of the fair sharing algorithm. $T_r$, $T_o$, $EV_r$, and $n$ represent transformer rate, transformer output, EV charging rate, and number of EVs connected to the transformer, respectively. If the charging rate is negative, the EVSE will stop charging. However, if the calculated charging rate is bigger than 7.2KWh, the EVSE adjusts the charging current to 30A.

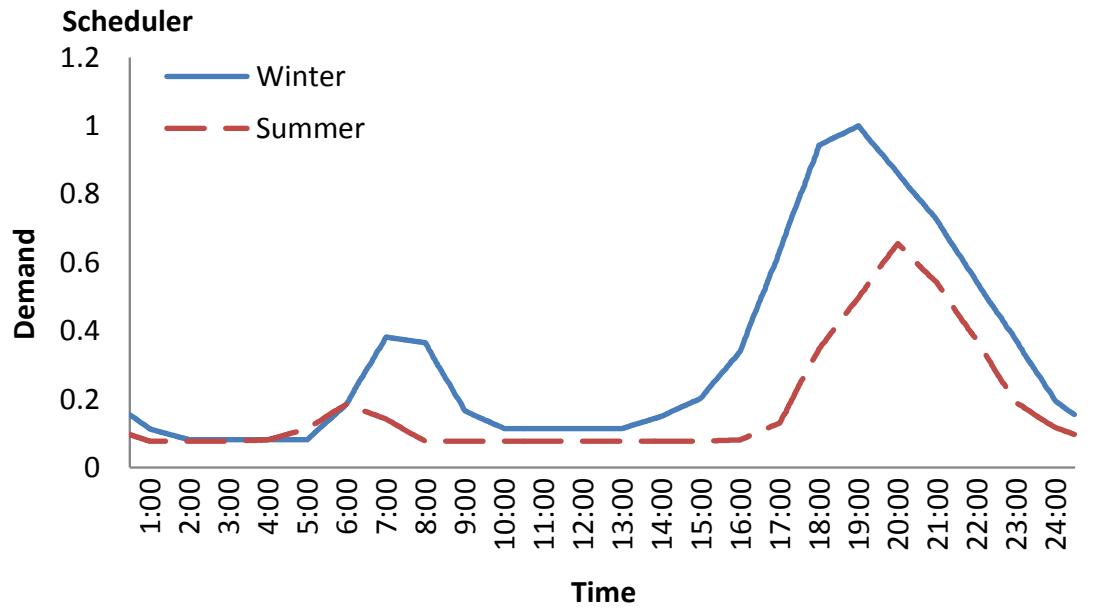

Fig. 4. Light Schedule

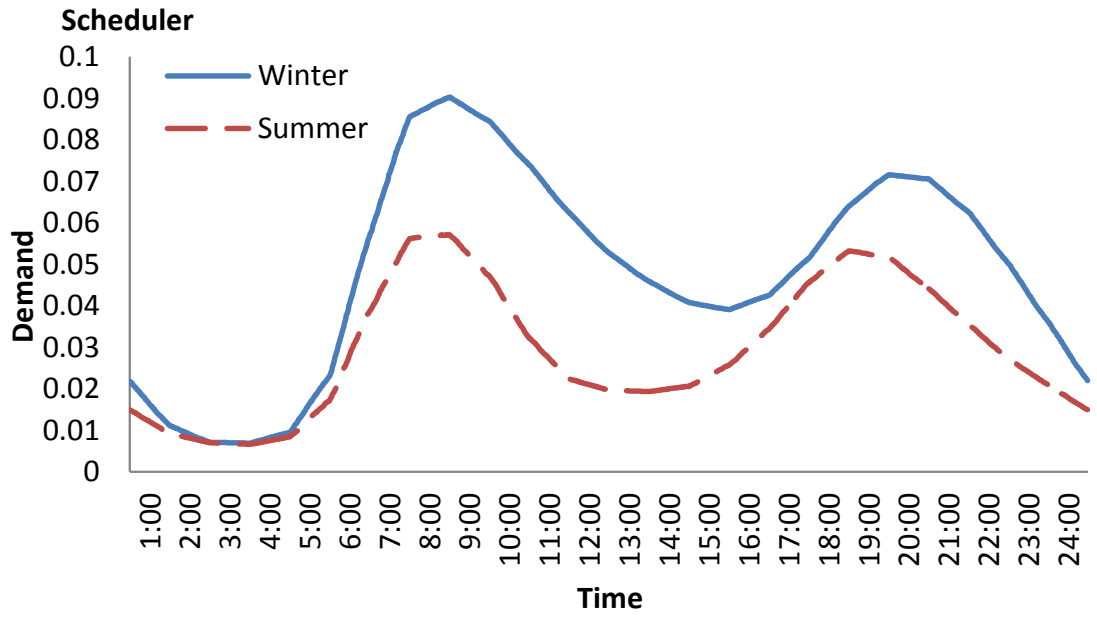

Fig. 5. Water Heater Schedule

# V. Simulation Result

We have simulated the power system model from January 3rd to 4th, 2012 for Winter. The simulation for a Summer day runs from August 2nd to August 3rd in the same year. We have compared the simulation results of our model with different sources [17][18]. In [11], we have validated our simulation results for the developed residential distribution grid model in greater detail.

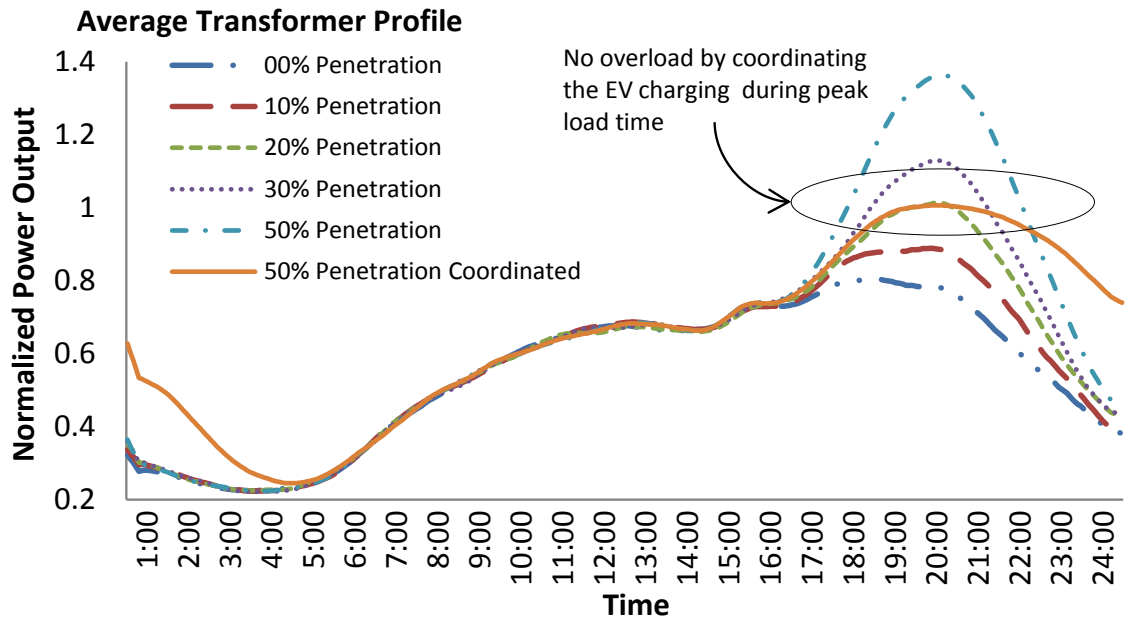

Fig. 6. Average Transformer-Level Power Output

We have simulated our residential distribution grid model according to five different EV penetration rates (0%, 10%, 20%, 30%, and 50%). For each penetration rate, we have randomly distributed the EVs among all houses; for example, 100 EVs were distributed among 1000 houses for a 10% penetration rate. Figure 6 shows the average power output for the step-down transformer level. For 30% and 50% penetration, the average normalized transformer output is above 1 during peak load time which means that in average transformers overloaded during this time.

[2] Our proposed algorithm is for advanced EVSEs which are able to provide variable rate current supply (multiple amperage adjustment capability) on-demand [19]

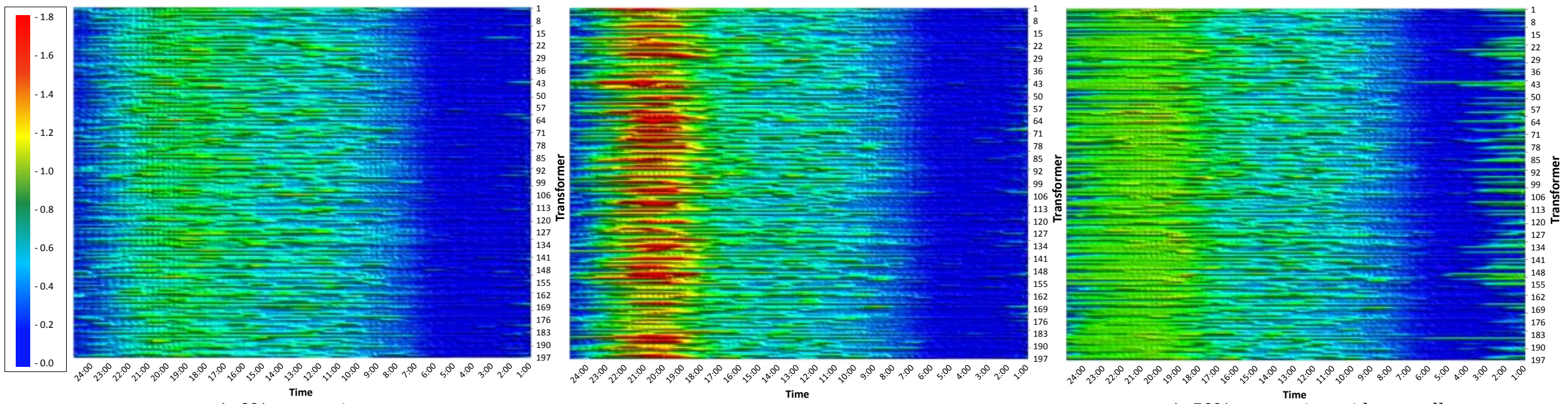

a) 0% penetration b) 50% penetration c) 50% penetration with controller

Fig. 7. Transformer-Level Power Output for Different EV Penetration Rate

However, the output power does not exceed the maximum rate of transformer by coordinating the EV charging. In Table III, we summarize the simulation results for different EV penetration rate. According to this table for 50% penetration rate, 60% of transformers are overloaded by EV charging and the maximum duration is about 5 hours. The aggregated overloading time is more than 200 hours for 50% penetration. The long overloading time will reduce the transformer life-time significantly. The EV charging effect can be mitigated by coordinating the EV charging. By increasing the penetration rate, the number of overloaded residential transformers increases exponentially during evening, when all the EVs arrive. Figure 7 (a) and (b) show the transformer level power output during a single day for 0% and 50% EV penetration rates. The vertical axis shows the transformers and the horizontal axis shows time. We normalized the output power of the transformer to its rate. The blue color shows low ratio of power output to transformer rate while the red color shows high ratio which means the transformer is overloaded heavily. In general, when the color changes to yellow (and consequently the red color), it means that the transformer must deliver much more power than its nominal rate. Figure 7 (c) illustrates the simulation result for coordinated EV charging (fair sharing algorithm).

TABLE III. TRANSFOMER OVERLOAD DATA

| EV Penetration | Overloaded Transformers | Maximum Duration (min) | Aggregated overloading time (min) |
|---|---|---|---|
| 0% | 0% | 0 | 0 |
| 10% | 3% | 137 | 489 |
| 20% | 10% | 179 | 1132 |
| 30% | 26% | 240 | 4744 |
| 50% | 60% | 327 | 13006 |
| 50% * | 0% | 0 | 0 |

*Collaborative and coordinated EV charging (fair sharing algorithm)

## VI. CONCLUSION

In this paper, we have modeled a residential distribution grid for different EV penetration rates. The EV model which we presented in this paper shows a high rate of overloading on step-down transformers (60% for 50% of penetration rate) during peak load time in residential distribution grid. However, by coordinating the EV charging in residential distribution grid, we may overcome transformer overloading without any upgrade in the power grid infrastructure.

## BIOGRAPHY

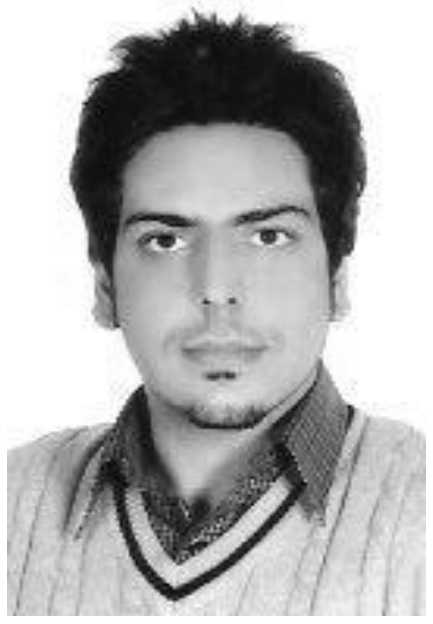

**Fereidoun Ahourai** is currently pursuing his Ph.D. degree in Electrical and Computer Science with a concentration in Computer System and Software at University of California, Irvine. He received the B.S. degree in Electrical Engineering from Zanjan University and the M.S. degree in Electrical Engineering with a concentration in Digital Electronics from Sharif University of Technology, Tehran, Iran. His research interest includes model-based design methods of cyber-physical energy systems, formal modeling of the residential microgrid, and demand-side energy management on the microgrid.

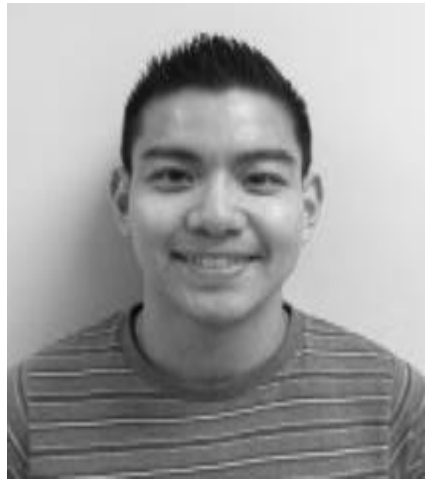

**Irvin Huang** is a fifth year electrical engineering major at the University of California, Irvine who has previously conducted research on the biomechanics of insects. He is now interested in the smart grid and the upcoming research needed to support its development. He is actively involved in his school's IEEE student branch and will continue to support IEEE as the GOLD (Graduates Of the Last Decade) coordinator for the Orange County area.

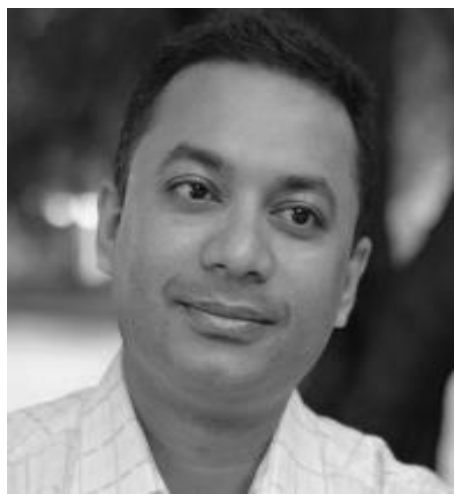

**Mohammad Abdullah Al Faruque** is currently with University of California Irvine (UCI), where he is a tenure track assistant professor and directing the Emulex Career Development Chair and the Advanced Integrated Cyber-Physical Systems (AICPS) lab. He is also affiliated with the Center for Embedded Computer Systems (CECS) at UCI. Before joining at UCI, he was with Siemens Corporate Research and Technology at Princeton, NJ as a research scientist. His current research is focused on system-level design of embedded systems and/or Cyber-Physical-Systems (CPS) with special interest on systems modeling, real-time scheduling algorithms, multi-core systems, dependable CPS systems, etc.